\documentclass[11pt]{article} \usepackage{axodraw} \usepackage{epsfig}
\newcommand{\la}[1]{\label{#1}}
\newcommand{\be}{\begin{equation}}
\newcommand{\ee}{\end{equation}}
\newcommand{\ba}{\begin{eqnarray}}
\newcommand{\ea}{\end{eqnarray}}
\newcommand{\bi}{\begin{itemize}}
\newcommand{\ei}{\end{itemize}}
\newcommand{\rmi}[1]{{\mbox{\scriptsize #1}}}
\newcommand{\nr}[1]{(\ref{#1})}
\newcommand{\tr}{{\rm Tr\,}}
\newcommand{\re}{\mathop{\rm Re}}

\newcommand{\fr}[2]{{\frac{#1}{#2}}}

\renewcommand{\vec}[1]{{\bf #1}}

\newcommand{\RR}{{\rm I\kern -.2em  R}}
\newcommand{\eq}{Eq.~}
\newcommand{\eqs}{Eqs.~}
\newcommand{\fig}{Fig.~}

\newcommand{\se}{Sec.~}
\newcommand{\ses}{Secs.~}

\def\lsi{\raise0.3ex\hbox{$<$\kern-0.75em\raise-1.1ex\hbox{$\sim$}}}
\def\gsi{\raise0.3ex\hbox{$>$\kern-0.75em\raise-1.1ex\hbox{$\sim$}}}
\newcommand{\lsim}{\mathop{\lsi}}

\makeatletter \@addtoreset{equation}{section} \makeatother
\renewcommand{\theequation}{\arabic{section}.\arabic{equation}}
\makeatletter
\renewcommand\section{\@startsection {section}{1}{\z@}%
                                   {-5.5ex \@plus -1ex \@minus -.2ex}% bfr-skip
                                   {2.3ex \@plus.2ex}%
                                   {\normalfont\large\bfseries}}
\renewcommand\subsection{\@startsection{subsection}{2}{\z@}%
                                     {-3.25ex\@plus -1ex \@minus -.2ex}%
                                     {1.5ex \@plus .2ex}%
                                     {\normalfont\normalsize\bfseries}}
\renewcommand\thesection {\@arabic\c@section}
\renewcommand\thesubsection   {\thesection.\@arabic\c@subsection}
\renewcommand{\@seccntformat}[1]{%
\csname the#1\endcsname.\hspace{1.0em}}
\makeatother

\begin{document}

\begin{titlepage}
\begin{flushright}
CERN-TH/2002-304\\
CPT-2002/P.4442\\
DESY-02-178 \\
IFIC/02-52 \\
FTUV-02-1111\\
hep-lat/0211020\\
\end{flushright}
\begin{centering}
\vfill
 
{\Large{\bf Finite-Size Scaling of Vector and Axial Current Correlators}}

\vspace{0.8cm}

P.H.~Damgaard$^{\rm a}$, 
P.~Hern\'andez$^{\rm b,}$\footnote{On leave
from Dept. de F\'{\i}sica Te\'orica, Universidad de Valencia.}, 
K.~Jansen$^{\rm c}$,
M.~Laine$^{\rm b}$, 
L.~Lellouch$^{\rm d}$

\vspace{0.8cm}

{\em $^{\rm a}$%
Niels Bohr Institute, Blegdamsvej 17, DK-2100 Copenhagen \O, Denmark\\}

\vspace{0.3cm}

{\em $^{\rm b}$%
Theory Division, CERN, CH-1211 Geneva 23, Switzerland\\}

\vspace{0.3cm}

{\em $^{\rm c}$%
NIC/DESY Zeuthen, Platanenallee 6, D-15738 Zeuthen, Germany\\}

\vspace{0.3cm}

{\em $^{\rm d}$%
Centre Physique Th\'eorique, CNRS, Case 907, Luminy,
F-13288 Marseille, France\\}

\vspace*{0.8cm}

{\bf Abstract}
 
\end{centering}
 
\vspace*{0.4cm}

\noindent
Using quenched chiral perturbation theory, we compute the
long-distance behaviour of two-point functions of flavour non-singlet
axial and vector currents in a finite volume, 
for small quark masses, and at a fixed gauge-field topology.
We also present the corresponding predictions for the unquenched 
theory at fixed topology. These results can in principle be used 
to measure the low-energy constants of the chiral Lagrangian, from 
lattice simulations in volumes much smaller than one pion Compton wavelength.
We show that quenching has a dramatic effect on the vector correlator, which 
is argued to vanish to all orders, while the axial correlator appears to be
a robust observable only moderately sensitive to quenching. 
\vfill
\noindent
 
%\noindent
%PACS numbers: 

%11.15.Ha, %        Lattice gauge theory
%11.30.Hv, %        Flavour symmetries
%11.30.Rd, %        Chiral symmetries
%12.38.Gc, %        Lattice QCD calculations
%12.39.Fe, %        Chiral Lagrangians
%\\
%Keywords:

\vspace*{1cm}
 
\noindent
CERN-TH/2002-304\\
November 2002  %% \today

\vfill

\end{titlepage}

%%%%%%%%%%%%%%%%%%%%%%%%%%%%% SECTION %%%%%%%%%%%%%%%%%%%%%%%%%%%%%%%%%%%
%
\section{Introduction}

The low-energy dynamics of QCD is governed by a spontaneously broken
chiral symmetry. Thus, close to the chiral limit, the correlation
length associated with the Goldstone bosons is very large. In
a finite volume, as required by lattice simulations, this correlation
length can easily overtake the linear extent of the box. When this
happens, the zero-momentum modes of the Goldstone bosons can
no longer be treated perturbatively and finite-size effects become
important. 

The general procedure for describing this situation in
chiral perturbation theory is the $\epsilon$-expansion of Gasser and
Leutwyler~\cite{GL} (see also~\cite{N}). It yields precise predictions
for the volume and quark-mass dependences of long-distance observables
in terms of a few infinite volume low-energy constants. The comparison
of these predictions with the volume and mass dependences of the same
observables computed in lattice QCD then permits the extraction
of the constants. Such lattice studies have become possible
recently thanks to the advances with Ginsparg--Wilson formulations \cite{GW}
of lattice fermions which implement a continuum-like chiral symmetry
at finite lattice spacing \cite{ML}.

In full QCD, all the relevant two-point functions of currents and
densities have already been calculated in the $\epsilon$-regime~\cite{H,HL}. 
However, due to the
numerical cost of Ginsparg--Wilson fermions, lattice simulations 
in this regime are currently restricted to the quenched
approximation. It is therefore important to evaluate the modifications
brought about by this approximation. In a previous paper~\cite{DDHJ}, 
scalar and
pseudoscalar correlators were calculated in the quenched approximation,
in sectors of fixed topology. We extend that work here by presenting
the calculation of flavour non-singlet axial and vector correlators.

At leading order in the chiral expansion, the chiral Lagrangian is
parametrized by two low-energy constants, the chiral condensate
$\Sigma$ and the pion decay constant $F$. The value of the
first one can be accurately extracted from the spectral 
density of the low-lying eigenvalues of the Dirac operator (see, e.g., 
ref.~\cite{eigenvalue}). It has also been obtained 
by measuring directly the quenched finite-volume quark 
condensate~\cite{DEHN}--\cite{Has}.  
To determine $F$, on the other hand,
we note that because it quantifies the strength of the
coupling of Goldstone boson fluctuations to the vacuum, it is natural
to consider correlation functions of these fluctuations. 
As shown in~\cite{H,HL}, the scalar and pseudo-scalar correlation
functions are proportional to $\Sigma^2$ at leading order, with $F$
appearing at the next-to-leading order. 
However, in the quenched approximation
these correlators also depend on additional unphysical
constants ($m_0^2$ and
$\alpha$), associated with the flavour singlet field~\cite{DDHJ}, 
which clearly makes numerical determinations rather
difficult, while maybe not impossible~\cite{PO}.

By contrast, the axial and vector current correlators are
proportional to $F^2$ at leading order, with $\Sigma$ appearing only
at the next order. In addition, these correlators are independent of
the unknown singlet constants $m_0^2$ and $\alpha$ at next-to-leading
order in the quenched theory. It is therefore expected that these
two-point functions will yield particularly clean determinations 
of (the quenched) $F$.

Our paper is organized as follows. In \se\ref{notation} 
we set up our notation and reiterate a few facts about the 
$\epsilon$-expansion of chiral perturbation theory. In \ses\ref{results} 
and \ref{nrts} we present our results for flavoured 
vector and axial current correlators in the quenched theory. Finally, 
we present in \se\ref{full} the analogous results for the full 
theory, and we conclude in \se\ref{conclu}. 

%%%%%%%%%%%%%%%%%%%%%%%%%%%%%%% SECTION %%%%%%%%%%%%%%%%%%%%%%%%%%%%%%%%%%%%%
\section{Basic setup}
\label{notation}

Let us consider QCD in a toroidal volume $V$ of average length scale
$L = V^{1/4}$. We assume that the volume is large with respect to the 
QCD scale, i.e. $F L \gg 1$.
As in infinite volume the lightest degrees of 
freedom are the would-be Goldstone
bosons of chiral symmetry breaking. Their interactions can be described in 
terms of a chiral Lagrangian, which can be systematically expanded in 
powers of the pion momentum and mass
 over the cutoff of the effective theory, 
$\Lambda \simeq 4 \pi F$. To leading order in the chiral expansion, 
\be
 {\cal L}_{\chi PT} = 
 \frac{F^2}{4} \tr \Bigl[\partial_\mu U\partial_\mu U^\dagger \Bigr] - 
 \frac{\Sigma}{2} \tr \Bigl[ 
 U M e^{i\theta/N_f} + M^\dagger U^\dagger e^{-i\theta/N_f}
 \Bigr], 
 \la{LE}
\ee
where $U\in $ SU($N_f$), $\theta$ is the vacuum angle, and, 
to this order,  $F$ and $\Sigma$
equal the pseudoscalar decay constant and
the chiral condensate, respectively. 
For simplicity, we consider the quark mass matrix $M$ 
to be diagonal, $M=\mathop{\mbox{diag}}(m,\ldots,m)$.

In an infinite volume both the momentum and the pion mass 
are taken to be of the same order. For a finite $L$ smaller
than the inverse of the pion mass, this conventional chiral 
perturbation theory breaks down when $m \Sigma V \lsim 1$, 
because the momentum zero-modes of the pion 
fields become non-perturbative~\cite{GL} and require a resummation to 
all orders. Gasser and Leutwyler provided a procedure to rearrange 
the chiral expansion appropriately~\cite{GL}. 
The first step is to factorise the zero and non-zero modes by writing
\be
 U = \exp\left(i \frac{2\xi}{F}\right) U_0, \la{U_0} 
\ee
where $U_0$ is a constant (in space-time) matrix of SU($N_f$), and the fields
$\xi(x)$ parametrize the non-zero-mode manifold. The power counting rules are 
then those of the so-called $\epsilon$-expansion:
\be
 F \sim {\cal O}(1), \quad
 \partial_\mu \sim {\cal O}(\epsilon), \quad
 L \sim {\cal O}(1/\epsilon), \quad
 \xi \sim {\cal O}(\epsilon), \quad
 m \sim {\cal O}(\epsilon^4). \la{epsexp}
\ee 
Thus, the quark mass goes like four powers of the momenta, instead of two. 

The difficulty of the $\epsilon$-expansion is simply that the integration 
over $U_0$ has to be carried out exactly.  
Such integrations can however be performed, at least numerically, 
and results for meson correlators have been computed in full QCD at 
next-to-leading order in the $\epsilon$-expansion~\cite{H,HL}. 
In some cases, it might be interesting to consider these observables 
also in sectors of fixed gauge-field topology (particularly when small 
physical volumes are used). The fixing of topology 
in chiral perturbation theory amounts to an enlargement of the zero-mode 
manifold from SU($N_f$) to U($N_f$), and to the addition of the term 
$(\det U_0)^\nu$ to the weight in the zero-mode integral~\cite{LS}, but 
then the integration can even be carried out analytically. 
 
Unfortunately, lattice simulations with very light quarks are still 
restricted to the quenched approximation. Therefore, analytical
predictions will also have to be obtained using quenched chiral 
perturbation theory. 

Two recipes have been proposed for dealing with the quenched limit. 
One is the so-called supersymmetric quenched chiral perturbation 
theory~\cite{BG,S}, where the cancellation of internal fermion loops 
is achieved by adding to the theory ghost bosons of spin $1/2$, one 
for each fermion field. A flavour singlet meson field, $\Phi_0$, also
needs to be added. The other is the so-called replica method \cite{DS}, 
in which only the flavour singlet field $\Phi_0$ (whose
non-zero momentum modes then correspond 
to $\tr \xi$ in \eq\nr{U_0}) is added, so that the manifold is 
U($N_f$). Computations are carried out by keeping 
track separately of the $N_v$ external flavours appearing in the 
operators and of the $N_f$ propagating dynamical flavours, and the quenched 
predictions are obtained by taking
the limit $N_f \rightarrow 0$, for a fixed $N_v$. 

While it is believed that the two methods are equivalent in perturbation
theory, the situation is somewhat trickier in the $\epsilon$-regime, 
where a non-perturbative definition of the integration over the 
zero-mode manifold is required. In particular, the replica 
method has so far provided results for the zero-mode integrals only
in terms of series expansions in a mass parameter
(see, however, the recent developments in~\cite{KSV}). Therefore, 
replica computations are usually transformed into  
supersymmetric ones at this point.

In the supersymmetric formulation, the zero-mode manifold is obtained
by enlarging U($N_f$) to a graded group. The simplest choice is to take 
U($N_f|N_f$) with $N_f = N_v$, although in principle any other choice 
with a larger $N_f \ge N_v$ should be equivalent. 
To leading order in the momentum expansion 
the supersymmetric quenched 
chiral Lagrangian is
\ba
 {\cal L}_{Q\chi PT} \!\! & = & \!\! \frac{F^2}{4} {\rm Str } 
 \Bigl[ \partial_\mu U \partial_\mu U^{-1} \Bigr] 
 - {m \Sigma \over 2} {\rm Str}
 \! \Bigl[ U_{\theta} U + U^{-1} U^{-1}_{\theta}\Bigr] %% \nn 
 %% & + & 
 + {m_0^2 \over 2 N_c} \Phi^2_0 + \frac{\alpha}{2N_{c}}
 (\partial_{\mu} \Phi_0)^2, \hspace*{0.5cm} %% \partial_{\mu}\Phi_0,
 \label{Lrep}
\ea
where $\rm Str$ denotes the supertrace, 
$\Phi_0 \equiv \frac{F}{{2}} 
{\rm Str}[-i \ln(U)]$ and $U_\theta\equiv
\exp(i \theta {I_{N_v}} /N_v)$. Here, ${I_{N_v}}$
is the identity matrix in the fermion--fermion block of ``physical''
Goldstone bosons and zero otherwise. 
At this order there are again the two couplings $\Sigma$ and $F$,
as in~\eq\nr{LE}, but also 
a series of new couplings, $m_0^2, \alpha,...$,  
associated with the flavour singlet field, which cannot 
be decoupled in the quenched limit~\cite{BG,S}. 

As discussed in \cite{D01,DDHJ}, 
the $\epsilon$-expansion can be set up for this 
theory as well,  at a fixed topological charge $\nu$.
The integral over the zero-momentum modes requires, however, 
special care. To obtain convergent answers, the manifold is 
in fact not taken to be U($N_v|N_v$) but what in the
mathematics literature is called the maximally symmetric 
Riemannian submanifold $\widehat{\rm Gl}(N_v|N_v)$ \cite{Z}. 
The resulting quenched integrals are known explicitly
for $N_v=1$~\cite{OTV,DOTV}, and some of them also
for $N_v=2$ \cite{TV}. 

In ref. \cite{DDHJ} the scalar and 
pseudoscalar two-point functions were computed in the 
$\epsilon$-expansion to next-to-leading order. 
Below we determine the analogous
vector and axial vector correlators in this expansion, at the
same relative order. The reader can find  
further technical details on the method in~\cite{DDHJ}. 

%%%%%%%%%%%%%%%%%%%%%%%%%%%%%% SECTION %%%%%%%%%%%%%%%%%%%%%%%%%%%%%%%%%%%%
\section{Quenched vector and axial current correlators}
\label{results}

To define the observables considered at the quark level, 
consider for definiteness a convention where the 
Euclidean QCD Lagrangian is ${\cal L}_E = \bar\psi \gamma_\mu D_\mu \psi$, 
with $\gamma_\mu^\dagger = \gamma_\mu$, 
$\{\gamma_\mu,\gamma_\nu\} = 2 \delta_{\mu\nu}$, and
$\gamma_5 = \gamma_0 \gamma_1 \gamma_2 \gamma_3$.  
The vector and axial currents can then be chosen as
\ba
 V^a_\mu(x) ~\equiv~ i \bar{\psi}(x) T_{N_v}^a \gamma_\mu \psi(x)~,~~~~~~~
 A^a_\mu(x) ~\equiv~ i \bar{\psi}(x) T_{N_v}^a \gamma_\mu \gamma_5 \psi(x) ~,
 \label{currents}
\ea
where $T^a_{N_v}$ are traceless generators in the physical flavour space of 
the $N_v$ valence quarks. We will work with the conventional normalization,
\be
 {\rm Tr} [ T_{N_v}^a T_{N_v}^b ] = \frac{1}{2} \delta^{ab}.
 \label{norm}
\ee
The corresponding currents in the chiral theory can be 
obtained by coupling covariantly the pion field to the external vector 
$v_\mu^a$ and axial-vector $a_\mu^a$ fields, and then 
taking functional derivatives with respect to those sources. 
More precisely, the above conventions imply that 
the partial derivatives in the chiral 
theory are promoted to covariant ones as
\ba
\partial_\mu U \rightarrow \partial_\mu U + 
i (v^a_\mu - a^a_\mu) T_{N_v}^a U - i U T_{N_v}^a (v^a_\mu + a^a_\mu),
\ea
and the currents become
\ba
 \Bigl(V_\mu^a\Bigr)_\rmi{$\chi$PT} \equiv {\cal V}_\mu^a \equiv
 \left. \Bigl( \frac{\partial {\cal L}_{\chi PT}}{\partial v^a_\mu} \Bigr)
 \right|_{v^a_\mu = 0}  & = & -i \frac{F^2}{2} {\rm Tr}\left[T_{N_v}^a \Bigl(
 \partial_\mu U U^{-1} + \partial_\mu U^{-1} U \Bigr) \right], 
 \la{cJmua1} \\
 \Bigl(A_\mu^a\Bigr)_\rmi{$\chi$PT} \equiv 
 {\cal A}_\mu^a \equiv
 \left. \Bigl( \frac{\partial {\cal L}_{\chi PT}}{\partial a^a_\mu} \Bigr)
 \right|_{a^a_\mu = 0} 
 & = & -i \frac{F^2}{2} {\rm Tr}\left[T_{N_v}^a \Bigl(
 -\partial_\mu U U^{-1} + \partial_\mu U^{-1} U \Bigr)\right].
 \la{cJmua2}
\ea

Inserting here \eq\nr{U_0}, expanding in $\xi$, carrying out 
the contractions, and taking the replica limit $N_f \to 0$, 
the two-point functions for ${\cal V}^a_\mu(x), {\cal A}^a_\mu(x)$
are easily determined at next-to-leading order. 
The general result can be written in 
a compact form by introducing  
$t^a_\pm \equiv T_{N_v}^a \pm U_0 T_{N_v}^a U_0^{-1}$,
and denoting ${\cal O}^{a,-}_\mu(x) \equiv {\cal V}^{a}_\mu(x)$, 
${\cal O}^{a,+}_\mu(x) \equiv {\cal A}^{a}_\mu(x)$, whereby 
(omitting contact terms)
\ba
 \bigl\langle {\cal O}^{a,\sigma}_\mu(x)~ 
 {\cal O}^{b,\tau}_\zeta(0) \bigr\rangle_\nu 
 & = &  - \frac{F^2}{2} 
 \sigma\tau \bigl\langle {\rm Tr} [t^a_{\sigma} t^b_{\tau}]
 \bigr\rangle_{\nu,U_0}^{\mu'} \; {\partial_{\mu}}{\partial_{\zeta}} 
 \bar{\Delta}(x-0)  \nonumber\\
 & & \hspace*{-2cm} - \frac{m\Sigma}{4} \sigma\tau
 \bigl\langle {\rm Tr}[\{ t^a_{\sigma}, t^b_{\tau}\}(U_0+ U^{-1}_0)] 
 \bigr\rangle_{\nu,U_0}^{\mu}  
 \int {\rm d}^4 z~\partial_{\mu} \bar{\Delta}(z-x) 
 \partial_{\zeta} \bar{\Delta}(z-0) \;. \label{res1} \label{res2}
\ea
%% \ba
%%  \left\langle {\cal V}^a_\mu(x)~ {\cal V}^a_\mu(0) \right\rangle_\nu 
%%  & = &  - \frac{F^2}{2} \left\langle {\rm Tr}\left[(t^a_-)^2
%%  \right]\right\rangle_{\nu,U_0} \;\left({\partial^x_{\mu}}\right)^2 
%%  \bar{\Delta}(x-0)  \nonumber\\
%%  & &  - \frac{m\Sigma}{2} \langle {\rm Tr}[(t_-^a)^2 (U_0+ U^{-1}_0)] 
%%  \rangle_{\nu,U_0}  \int d^4 z~\partial^z_{\mu} \bar{\Delta}(x-z) 
%%  \partial^z_{\mu} \bar{\Delta}(0-z) \;, \label{res1}\\
%%  \left\langle {\cal A}^a_\mu(x)~ {\cal A}^a_\mu(0) \right\rangle_\nu 
%%  & = & - \frac{F^2}{2} \langle {\rm Tr}[(t_+^a)^2]\rangle_{\nu,U_0} 
%%  \;\left({\partial^x_{\mu}}\right)^2 \bar{\Delta}(x-0)  \nonumber\\
%%  & & - \frac{m\Sigma}{2} \langle {\rm Tr}[(t_+^a)^2 (U_0+ U^{-1}_0)]
%% \rangle_{\nu,U_0} \int d^4 z~\partial^z_\mu \bar{\Delta}(x-z) 
%%  \partial^z_\mu \bar{\Delta}(0-z) \;, \hspace*{0.5cm}
%%  \label{res2}
%% \ea
Here
$\langle\cdots \rangle_{\nu,U_0}^{\mu}$ denotes the zero-mode average,
\be
 \langle \cdots \rangle_{\nu,U_0}^{\mu} ~\equiv~
 \frac{\int_{U_0} (\cdots) (\det U_0)^\nu \exp( \mu \re \tr[U_0])}
 {\int_{U_0} (\det U_0)^\nu \exp( \mu \re \tr[U_0])},   \la{zm}
\ee
while $\bar{\Delta}(x)$ is the 
meson propagator without zero-momentum modes,
\ba
{\bar \Delta}(x) \equiv \frac{1}{V} \sum_{p\neq 0} 
\frac{e^{i p \cdot x}}{p^2}. 
\ea
Furthermore, $\mu \equiv m \Sigma V$, while $\mu'$
is an order-$\epsilon^2$ corrected value thereof~\cite{DDHJ}; 
in practice, however, $\mu'$ is not needed here, 
since the leading order results 
turn out to be pure numbers.
The first term in \eq\nr{res1} is the leading, 
the second the next-to-leading order contribution. 
In the supersymmetric formulation the expression 
is formally identical, after the substitution 
${\rm Tr} \rightarrow {\rm Str}$ and the corresponding change of the 
group manifold for $U_0$.  

Before we go on to discuss the zero-mode integrations, let us emphasize two 
important points. First, none of the higher-order couplings of the
chiral Lagrangian (the $L_i$'s of Gasser and Leutwyler) 
contribute at this order.
This is in contrast to the standard $p$-expansion of chiral perturbation 
theory where, for example, $L_4$ contributes to the axial correlator at 
next-to-leading order~\cite{H,HL}. This shows the advantage of the 
$\epsilon$-regime
with respect to the $p$-regime in extracting the low energy constants
$\Sigma$ and $F$. Second, as shown by a comparison of \eq\nr{res1} with the
unquenched results (in \se\ref{full}), quenching does not modify 
the space-time dependence of these correlators because, remarkably, 
all potential double-pole contributions to the propagators 
cancel at both leading and next-to-leading orders.
In other words, the additional constants $m_0^2$ and 
$\alpha$ of the quenched theory do not appear at all. This is in contrast 
to the scalar and pseudoscalar correlators, which show more important 
quenching effects~\cite{DDHJ}. 
%These quenching artifacts are
%responsible for a rather poor convergence of the perturbative expansion. 
This implies that the extraction of $\Sigma$ and particularly $F$,
which appears at the first order,  
should be easier from the axial and vector currents. 
 
We now return to the zero-mode integrations in~\eq\nr{res1}.
As explained above, the replica method does not yet
provide a recipe to compute them, apart from series 
expansions. However, since the perturbative rules for the two
formulations agree, we can at this point switch to the supersymmetric
formulation. For a diagonal mass matrix, the results 
for the zero-mode integrations
are invariant in the linear transformation
$T^a_{N_v} \to g T^a_{N_v} g^{-1}, T^b_{N_v} \to g T^b_{N_v} g^{-1}$ 
for any $g \in \widehat{\rm Gl}(N_v|N_v)$, and should thus be 
proportional to the singlet tensor $\delta^{ab}$. 
For $\langle {\cal V}^a_\mu(x) {\cal V}^b_\zeta(0) \rangle$
and $\langle {\cal A}^a_\mu(x) {\cal A}^b_\zeta(0) \rangle$, 
it is then sufficient to consider
\ba
 {\cal I}^0_{\pm}\equiv \bigl\langle 
 {\rm Str}\bigl[\left(t_{\pm}^a\right)^2 \bigr] 
 \bigr\rangle_{\nu,U_0}^{\mu'}\;,\;\;\; 
 {\cal I}^1_{\pm}\equiv\bigl\langle {\rm Str}
 \bigl[\left(t_{\pm}^a\right)^2 (U_0+U_0^{-1} )\bigr]
 \bigr\rangle_{\nu,U_0}^{\mu} \;, \label{i0andi1}
\ea
where the superscript $a$ is not summed over. The superscripts in 
the ${\cal I}$'s, on the other hand,  
refer to the order in the expansion at which these 
contributions enter, as mentioned above.

It is believed that we can now take, 
without loss of generality, the generators $T^a_{N_v}$ 
to be in the simplest group allowing to define flavour non-singlets, 
i.e. $N_v =2$. 
The first integrals in~\eq\nr{i0andi1} can then be evaluated 
by a series of tricks explained in ref.~\cite{DDHJ}. 
Using the explicit integrations over $\widehat{\rm Gl}(2|2)$ 
obtained in \cite{TV}, it is easy to show that for 
the normalization of \eq\nr{norm},
\be
 {\cal I}^0_- = 0 \;,\;\;\;\;\;\;\; {\cal I}^0_+ = 2 \;.\;\;\; 
 \label{magic0}
\ee
{}The striking result is that the vector correlator 
vanishes identically at leading order. 

The integrals at the next order are not as easy to obtain, because third
derivatives of the partition function for $N_v = 2$ are needed, and they 
have to our knowledge not been computed before. We have 
however performed large 
and small mass expansions  in the replica method
using the results of \cite{DS1}, 
and found that ${\cal I}^1_-$ vanishes to leading order in both of 
these expansions. 
(This, in turn, would imply that 
${\cal I}^1_+ = (2/N_v) \langle \tr [U_0 + U_0^{-1}]\rangle_{\nu,U_0}^{\mu}$.) 
This can hardly be a coincidence: it suggests
that also ${\cal I}^1_-$ is identically zero. It should be noted that 
in the unquenched theory the non-singlet vector 
correlator is certainly non-vanishing~\cite{H,HL}.

In order to understand this feature of the 
quenched approximation better, we present in the next section 
a new way of obtaining these results.  
In fact, we will argue that the vector correlator 
vanishes in the quenched approximation to all orders in
chiral perturbation theory, both in the $p$- and 
$\epsilon$-regimes. The same insight provides a useful trick 
to obtain certain integrals in $\widehat{\rm Gl}(N_v|N_v)$ from 
those based on the much simpler $\widehat{\rm Gl}(1|1)$. 

%%%%%%%%%%%%%%%%%%%%%%%%%%%%% SECTION %%%%%%%%%%%%%%%%%%%%%%%%%%%%%%%%%%%%%
\section{Quenched zero-mode integrals and an all-orders argument}
\label{nrts}

%To make the argument, we again use the language of the 
%replica method, in the sense of considering the limit $N_f \to 0$. 

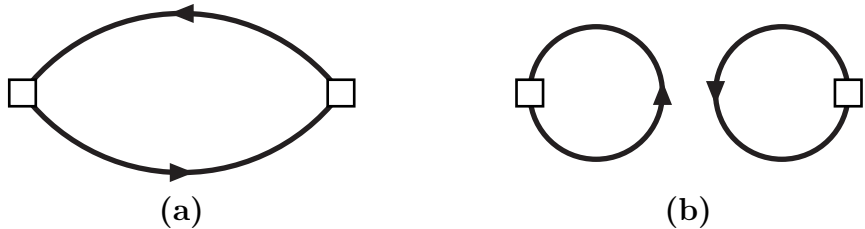
\begin{figure}[t]

%%
%% Original concept for axodraw macros: (C) York Schroder, 2001. 
%%
%% thick line
%%
\def\TAsc(#1,#2)(#3,#4,#5)%
{\SetWidth{2.0}\ArrowArc(#1,#2)(#3,#4,#5)\SetWidth{1.0}}
%%
%% a box 
%%
\newcommand{\pic}[1]{\;\parbox[c]{120pt}{\begin{picture}(120,60)(0,-20)
\SetWidth{1.0}\SetScale{1.0} #1 \end{picture}}\;}
%%
%% topologies for graphs (with line styles as arguments)
%%
\def\Connected(#1,#2){\pic{#1(60,-15)(75,34,146) #2(60,75)(75,214,326)%
\GBoxc(0,30)(10,10){1} \GBoxc(120,30)(10,10){1}% 
\Text(60,-15)[c]{\large\bf (a)} }}
\def\Disconnected(#1,#2){\pic{#1(25,30)(25,183,177) #2(95,30)(25,0,360)%
\GBoxc(0,30)(10,10){1} \GBoxc(120,30)(10,10){1}%
\Text(60,-15)[c]{\large\bf (b)} }}
%% 
%% the figure can then be conveniently done in a math environment
%%

\begin{eqnarray*}
 & & \hspace*{-1cm}
 \Connected(\TAsc,\TAsc) \qquad \qquad \qquad
 \Disconnected(\TAsc,\TAsc) 
\end{eqnarray*}

\caption[a]{\it Two types of external flavour contractions of two quark 
bilinears: (a) ``connected'' and (b) ``disconnected''.}  
\label{fig:nrt}

\end{figure}

The argument relies on a relation that can be derived at the quark level,
where the effect of quenching is immediately obvious.
We start by considering the two-point function of any 
flavour singlet quark bilinear, ${\cal O}^0 (x)$. 
In general there are two ways of contracting the external flavours 
leading to two types of diagrams, shown
in \fig\ref{fig:nrt}. In full QCD with $N_f$ flavours the first one, 
which we shall call ``connected'', will behave in terms of
$N_f$-counting as $N_f + {\cal O}(N_f^2)$, while
the second behaves as $N_f^2 + {\cal O}(N_f^3)$. 
The idea is now to isolate the quenched ``connected'' contribution 
of one external flavour ($N_v = 1$) by 
appropriately normalizing the approach 
of the full singlet quantity to the replica limit $N_f \to 0$, 
\be
 \langle {\cal O}^0(x) {\cal O}^0(y) 
 \rangle^\rmi{quenched, $N_v=1$}_\rmi{connected} 
 = \lim_{N_f\rightarrow 0} \frac{1}{N_f} \langle {\cal O}^0(x) {\cal O}^0(y) 
 \rangle^\rmi{full}. \label{nrt}
\ee  
The contribution of the ``disconnected'' diagram vanishes in this limit 
since it is $O(N_f^2)$. 

%%%%%%%%%%%%%%%%%%%%%%%%%% FIGURE %%%%%%%%%%%%%%%%%%%%%%%%%%%%%%%%%%%%%%%
%%%%%%%%%%%%%%%%%%%%%%%%%%%%%%%%%%%%%%%%%%%%%%%%%%%%%%%%%%%%%%%%%%%%%%%%%%

Now that the quenched connected contribution for one flavour 
($N_v = 1$) is isolated, it is straightforward 
to get any non-singlet ($N_v \ge 2$)
quenched correlator in the limit of degenerate quark masses. 
Indeed, the non-singlet correlators only contain the connected 
diagram and this diagram gives 
the same contribution no matter whether the two 
lines carry the same flavour. In other words, 
the $N_v$ dependence of this diagram is a trivial 
overall factor, which can easily be provided: the non-singlet quenched 
correlator %%of the operators in \eq(\ref{currents}) 
is obtained 
by multiplying \eq(\ref{nrt}) with the factor in \eq(\ref{norm}),
\be
 \langle {\cal O}^a(x) {\cal O}^b(y) 
 \rangle^\rmi{quenched} 
 = {\rm Tr}\left[T_{N_v}^a T_{N_v}^b\right] \cdot
 \lim_{N_f\rightarrow 0} \frac{1}{N_f} \langle {\cal O}^0(x) {\cal O}^0(y) 
 \rangle^\rmi{full} \;. \label{nrt2}
\ee

Up to this point we argued at the quark level. However, since the relation 
of \eq(\ref{nrt2}) must be valid at any space-time separation,  
it must hold also in the effective chiral theory. 
The only subtlety involved in computing the matrix element on the right hand 
side of \eq(\ref{nrt2}) is that the chiral Lagrangian for full QCD must 
include the singlet pseudoscalar field, since the decoupling of the 
singlet does not commute with the limit $N_f \rightarrow 0$. 

Assuming that the relation of 
\eq(\ref{nrt2}) indeed holds in the effective 
theory, it is easy to show that 
{\em the vector correlator vanishes to all 
orders in the quenched chiral theory}\footnote{Obviously there are 
non-vanishing shorter-distance contributions.}. 
The reason is that, if we were to 
replace $T^a_{N_v} \rightarrow T^0_{N_f}\equiv {I}_{N_f}$  
in \eqs\nr{cJmua1}, \nr{cJmua2} and write $U = \exp(i \, 2 \xi/F)$, 
then ${\cal V}^0_\mu = 0$, 
while ${\cal A}^0_\mu = -2 F \, \partial_\mu \tr \xi$. 
In other words, the vector singlet (i.e. baryon number) current
vanishes identically\footnote{%
 This is true to all orders, not only at the level 
 shown in~\eqs\nr{cJmua1}, \nr{cJmua2}.}, 
as does then also the limit
of \eq(\ref{nrt2}).  On the other hand, the axial singlet current couples to
the flavour singlet field, which cannot be decoupled in the 
quenched limit, leading to a non-vanishing contribution
for the axial current. 

Since the non-singlet vector current has the quantum numbers 
of the $\rho$ meson, we note that the result mentioned 
is consistent with, and in some sense provides justification for, 
the common lore that the quenched $\rho$ ``does not decay''.
%% into the 
%% light degrees of freedom contained in quenched chiral perturbation theory.

Let us stress that 
according to the argument just presented, the zero-mode integrals 
in the case of degenerate quark masses can all be obtained from 
the simplest case $N_v=1$. In the supersymmetric formulation 
this implies a set of very non-trivial relations 
between integrals in $\widehat{\rm Gl}(N_v|N_v)$ for various $N_v$. 
In particular, we have checked that one
can reproduce  in this way
all the results for the non-singlet scalar 
and pseudoscalar correlators of~\cite{DDHJ}, and those in \eqs(\ref{magic0}), 
without the explicit integrals over $\widehat{\rm Gl}(2|2)$ of \cite{TV}: 
the only non-trivial integrals needed are in  $\widehat{\rm Gl}(1|1)$. 

Similarly, the quenched singlet correlator for any $N_v$ can be 
easily computed, using the singlet computed for $N_v=1$, together with the
quenched connected contribution obtained as in \eq(\ref{nrt}). 
The difference of these two quantities is the quenched disconnected
contribution. However, 
once the connected and disconnected contributions for 
the singlet with $N_v=1$ are isolated, the full result for any $N_v$ is 
simply $N_v \times $(connected) + $N_v^2 \times$(disconnected). From the 
point of view of the effective theory, this implies again non-trivial 
relations between certain integrals 
in $\widehat{\rm Gl}(N_v|N_v)$ and $\widehat{\rm Gl}(1|1)$, or the 
corresponding replica limits of unitary integrals. 
For completeness, we illustrate this point in~\ref{app:example}.

\vspace*{0.5cm}

Using the above observations 
we can easily compute the integrals ${\cal I}^1_{\pm}$ in~\eq\nr{i0andi1}, 
and thus simplify the expression in~\eq(\ref{res1}). %%, (\ref{res2}). 
To display the result, it is convenient to consider 
the zero components of the currents, and to project onto 
zero spatial momenta by the usual Fourier integration. 
Using the relations (up to contact terms)
\ba
 \int_{\vec x}  \partial^2_0 \bar{\Delta}(x-0)  & = & 
 \frac{1}{T} \;,  \\
 \int_{\vec x}  \int_z \partial_0 \bar{\Delta}(z-x) 
 \partial_0 \bar{\Delta}(z-0)  & = & 
 T \, h_1\left(\frac{x_0}{T}\right), 
\ea
where $T$ is the temporal extent of the box and 
$h_1(\tau) \equiv \fr12 \Bigl[\Bigl(|\tau| - 
\fr12\Bigr)^2 - \fr{1}{12} \Bigr]$~\cite{HasenL}, we finally obtain
(no sum over $a$):
\ba
 C^\rmi{quenched}_v(x_0)\equiv\int_{\vec x}\langle 
 {\cal V}^a_0(x)~ {\cal V}^a_0(0) \rangle_\nu 
 & = & 0,  \label{vector} \\
 C^\rmi{quenched}_a(x_0)\equiv\int_{\vec x} \langle 
 {\cal A}^a_0(x)~ {\cal A}^a_0(0) \rangle_\nu 
 & = & - \frac{F^2}{T}\biggl[ 1 + \frac{2\, m \Sigma_\nu(\mu) T^2}{F^2} 
 h_1\left(\frac{x_0}{T}\right) \biggr], 
 \label{axial}
\ea
where the only non-trivial zero-mode integral is the same as appears 
in the quark condensate obtained with $\widehat{\rm Gl}(1|1)$ \cite{DOTV}: 
\be
{\Sigma_\nu (\mu) \over \Sigma} ~\equiv~ 
\mu \Bigl[ I_\nu (\mu) K_\nu (\mu) + I_{\nu +1} (\mu)
K_{\nu -1}(\mu) \Bigr] +{\nu \over \mu}, 
\label{zerocon} 
\ee
where $I_\nu, K_\nu$ are Bessel functions.
%% written in terms of the scaling variable $\mu\equiv m\Sigma V$. 

To illustrate the results of \eq(\ref{axial}), we show in 
Fig.~\ref{fig:aa} the axial correlator normalized to the tree level value,
for a realistic lattice setup (of volume $V \equiv T L^3$). 
Note that the curvature
at zero quark mass limit is quite sizeable, for $\nu > 0$.

%%%%%%%%%%%%%%%%%%%%%%%%%%%% FIGURE %%%%%%%%%%%%%%%%%%%%%%%%%%%%%%%%%%%%%%%
\begin{figure}[t]

\begin{center}
\epsfig{file=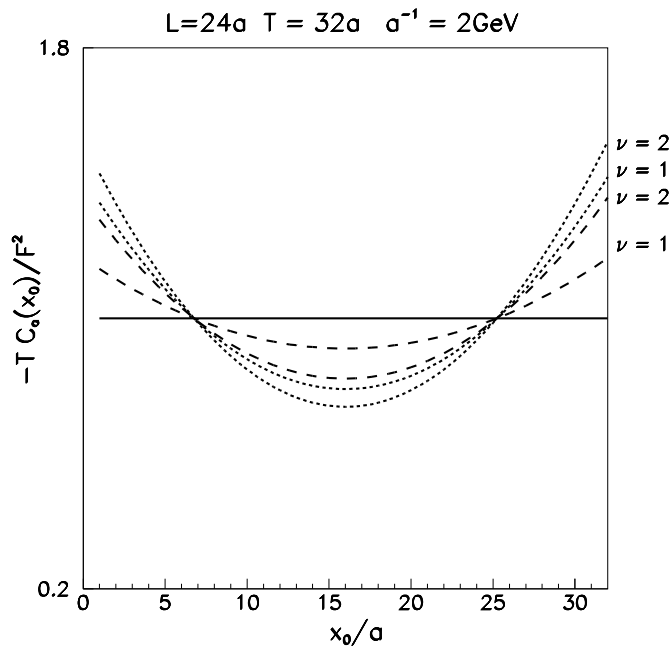,width=9cm}
\end{center}

\caption[a]{\it 
The expression inside the square brackets in 
\eq\nr{axial}, for $T = 32 a$, $L = 24 a$, $a^{-1} = 2$ GeV, and
two masses,  $m=0$ MeV (dashed) and $m=5$ MeV (dotted),
as well as a 
topological charge $\nu = 1,2$. The solid line is just the tree-level 
result. For the parameters in the chiral Lagrangian 
(which to this order are scale independent)
we have assumed $F = 93$ MeV, $\Sigma = (250 \mbox{ MeV})^3$. 
%% (Note that 
%% the scheme dependence of $m$ and $\Sigma$ cancels in the product that 
%% appears in the observables).
} 

\la{fig:aa}
\end{figure}
%%%%%%%%%%%%%%%%%%%%%%%%%%%%%%%%%%%%%%%%%%%%%%%%%%%%%%%%%%%%%%%%%%%%%%%%%%%

One apparent paradox
in \eqs(\ref{vector}), (\ref{axial}) is 
the following\footnote{We thank M.~L\"uscher for pointing out this paradox.}. 
By performing a  global chiral rotation, 
$\psi \to \psi' = \exp(i \phi^a T^a \gamma_5) \psi$, which should be
a good symmetry in the chiral limit at fixed volume, one can relate 
the vector and axial correlators through
\be
 \sum_a \langle V_\mu^a(x) V_\nu^a(0) \rangle = 
 \sum_a \langle A_\mu^a(x) A_\nu^a(0) \rangle - 
 \frac{1}{N_v} \sum_{a,b,c} f^{abc} \;
 \Bigl\langle 
 V^a_\mu(x) A^b_\nu(0) \delta_c S   
 \Bigr\rangle \;,
 \label{equal}
\ee
where $\delta_c S =  {{\rm d} S}/{{\rm d}\phi^c}$ is the 
variation of the action.
The second term on the right is clearly proportional to the bare
quark mass, since this is the only source of chiral symmetry breaking, 
and one might then expect that it vanishes in the chiral limit, 
leading to equal vector and axial correlators.

Clearly, \eqs(\ref{vector}) and (\ref{axial}) contradict this expectation.
To be explicit we have, in the limit of small quark masses, 
\be
 \Sigma_\nu(\mu) \rightarrow 
 \left\{ \begin{array}{lll} 
 &\Sigma \left({|\nu|\over \mu} + 
 {\mu \over 2 |\nu|} \right) + {\cal O}(\mu^3), & \nu\neq 0 \\
 &\Sigma \left(\frac{1}{2} - \gamma + \log\left(2/\mu\right) 
 \right) \mu  + {\cal O}(\mu^3), & \nu = 0
 \end{array} \right. \;,
\ee
which shows that the axial correlator in \eq(\ref{axial}) 
does not vanish in this limit. 
This is however not in contradiction with~\eq\nr{equal}, because
%% with the global symmetry identities, because when the averages in 
%% \eq(\ref{equal}) are taken in sectors of fixed topology, 
the second term on the right does not vanish either. The reason is that 
when averages are considered in sectors of fixed topology, there are chiral 
singularities in the three-point function 
on the right hand side of~\eq\nr{equal} that cancel 
the explicit factor of $m$, giving a finite
contribution in the chiral limit, which exactly matches the difference 
between the vector and axial correlators. While the situation for $\nu\neq 0$ 
is very similar in the full and quenched theories, and these singularities
can be clearly identified as the topological zero modes, in the case 
of $\nu=0$, the behaviours are different in the two cases: the vector 
and axial correlators are the same in the chiral limit in 
the full theory while they differ in the quenched case. 
This indicates a more singular chiral 
behaviour of the quenched theory, which was already noticed in the case
of the quark condensate \cite{DEHN,HJL}.

%%%%%%%%%%%%%%%%%%%%%%%%%%%%%%% SECTION %%%%%%%%%%%%%%%%%%%%%%%%%%%%%%%%%%%
\section{Results in the full theory}
\label{full}

For completeness and comparison we present in this section the results for 
the vector and axial correlators in the full theory at fixed topology. 
They can almost be read from \cite{H}, if not because there are a few places 
where identities were used for the zero-mode integrals that are not valid 
when averages are taken in sectors of fixed topology. After integrating 
over space and using the relations
\ba
 {\cal J}^1_- \equiv 
 \langle \tr [(t^a_{-})^2(U_0 + U_0^\dagger)] \rangle_{\nu,U_0}^{\mu}
 \!\! & = & \!\!
 \frac{N_f}{\mu}\bigl\langle \tr \bigl[ (t^a_{+})^2  %% \rangle_{\nu,U_0}
 - %% \langle \tr 
 (t^a_{-})^2\bigr] \bigr\rangle_{\nu,U_0}^{\mu} 
 \;, \\
 {\cal J}^1_+ \equiv 
 \langle \tr [(t^a_{+})^2(U_0 + U_0^\dagger)] \rangle_{\nu,U_0}^{\mu}
 \!\! & = & \!\! 
 \frac{N_f}{\mu}\bigl\langle \tr \bigl[ (t^a_{-})^2 %% \rangle_{\nu,U_0} 
  - %% \langle tr 
 (t^a_{+})^2 \bigr] \bigr\rangle_{\nu,U_0}^{\mu}  + 
 \frac{4}{N_f}\langle \re \tr U_0 \rangle_{\nu,U_0}^{\mu}
 , \hspace*{0.5cm}
\ea 
we obtain (no sum over $a$):
\ba
 C_v^\rmi{full}(x_0)\equiv\int_{\vec x}
 \langle {\cal V}^a_0(x) {\cal V}^a_0(0) \rangle_\nu 
 & = & -\frac{F^2}{2 T}\left\{{\cal J}^0_- 
 +\frac{N_f}{F^2} \left(\frac{\beta_1}{V^{1/2}} {\cal J}^0_- 
 -\frac{T^2}{V} k_{00} {\cal J}^0_+ \right) \right\}\;,\label{fullvector} \\
 C_a^\rmi{full}(x_0)\equiv\int_{\vec x}
 \langle {\cal A}^a_0(x) {\cal A}^a_0(0) \rangle_\nu 
 & = & -\frac{F^2}{2 T}\left\{{\cal J}^0_+ 
 +\frac{N_f}{F^2} \left(\frac{\beta_1}{V^{1/2}} {\cal J}^0_+ 
 -\frac{T^2}{V} k_{00} {\cal J}^0_- \right) \right.\nonumber\\
 & & \hspace*{1.75cm}\left. +\frac{4 \mu}{N_f F^2} 
 \frac{T^2}{V}h_1\Bigl(\frac{x_0}{T}\Bigr) 
 \langle \re \tr U_0 \rangle_{\nu,U_0}^{\mu}  \right\}\;,
 \hspace*{0.5cm}\label{fullaxial} 
\ea
where $\beta_1, k_{00}$ are numerical factors~\cite{HasenL,H}, and
\ba
 {\cal J}^0_{+} & \equiv & 
 \langle {\rm Tr}[(t^a_{+})^2] \rangle_{\nu,U_0}^{\mu'} 
 = \frac{1}{N_f^2-1} \left( N_f^2 - 2 +  
 \langle {\rm Tr}[U_0]\, {\rm Tr}[U_0^\dagger]
 \rangle_{\nu,U_0}^{\mu'} \right)\;, \la{J0p}\\
 {\cal J}^0_{-} & \equiv & 
 \langle {\rm Tr}[(t^a_{-})^2] \rangle_{\nu,U_0}^{\mu'} 
 = \frac{1}{N_f^2-1} \left( N_f^2 
 - \langle {\rm Tr}[U_0]\, {\rm Tr}[U_0^\dagger]\rangle_{\nu,U_0}^{\mu'}\right)
 \;. \la{J0m}
\ea
Here again $\mu = m \Sigma V$, while
its order-$\epsilon^2$ corrected value is~\cite{H,HL}
\be
 \mu' \equiv \mu \cdot \biggl( 
 1 + \frac{N_f^2 - 1}{N_f} \frac{\beta_1}{F^2 V^{1/2}} \biggr)\;. 
\ee
The average left over
in~\eqs\nr{J0p}, \nr{J0m} is given by~\cite{DDHJ}
\ba
 \langle {\rm Tr}[U_0] \,{\rm Tr}[U_0^\dagger]\rangle_{\nu,U_0}^{\mu'} 
 = N_f \left[ \frac{\Sigma_\nu'(\mu')}{\Sigma} 
 + N_f \left(\frac{\Sigma_\nu(\mu')}{\Sigma}\right)^2 
 + \frac{1}{\mu'} \frac{\Sigma_\nu(\mu')}{\Sigma} 
 - N_f \frac{\nu^2}{(\mu')^2} \right],
\ea
where
$\Sigma_\nu(\mu') = (\Sigma/N_f) \langle 
\re \tr U_0 \rangle_{\nu,U_0}^{\mu'}$ 
is the unquenched chiral condensate at fixed topology (see for 
instance~\cite{LS}). 
Note that the unquenched $\Sigma_\nu(\mu')/\Sigma$ 
depends also on $N_f$, although this has not been displayed explicitly. 

As we see from~\eq\nr{fullvector}, the vector charge is conserved
in the full theory just as in the quenched one, but it has now
a non-vanishing value. It can be easily checked, however, that 
at, say, $\nu\neq 0$ but $\mu\to 0$, one gets
  ${\cal J}^0_- = 0 + {\cal O}(N_f^2)$, 
  ${\cal J}^0_+ = 2 + {\cal O}(N_f^2)$,
  ${\cal J}^1_- = 0 + {\cal O}(N_f)$,
  ${\cal J}^1_+ = (2/N_f) \langle \tr [U_0 + U_0^\dagger]\rangle 
  + {\cal O}(N_f)$, 
in accordance with the quenched results in~\se\ref{nrts}
for ${\cal I}^0_-, {\cal I}^0_+, {\cal I}^1_-, {\cal I}^1_+$, 
respectively. 
In the full theory the vector correlator has also been computed 
at next-to-leading order in the $p$-expansion~\cite{H,HL} and, being 
proportional to $N_f$, again vanishes in the replica limit $N_f \to 0$.

%%%%%%%%%%%%%%%%%%%%%%%%%% SECTION %%%%%%%%%%%%%%%%%%%%%%%%%%%%%%%%%%%%%%%
\section{Conclusions}
\label{conclu}

We have presented the results for flavoured vector and axial vector 
two-point functions in quenched chiral perturbation theory, at next-to-leading 
order in the $\epsilon$-expansion, corresponding to 
a finite volume and the vicinity of the chiral limit. Analogous 
results for unquenched QCD were already presented in~\cite{H,HL}. 

We find that quenching has a striking effect on the vector 
correlator, which can be argued to vanish to all orders
in the chiral expansion. The axial correlator on the other hand depends
at the leading order on the low energy constant $F$, and 
at the next-to-leading order also on $\Sigma$. However, 
it does not depend to this order on any of the couplings 
of the quenched theory associated with the singlet pseudoscalar field, 
$m_0^2, \alpha,...,$ nor on the $L_i$'s of Gasser and Leutwyler. 
The measurement of the volume and mass dependence 
of this correlator close to the chiral limit would therefore 
permit the extraction of (the quenched) $\Sigma$ and $F$, 
with a minimal contamination from higher order effects. 

%%%%%%%%%%%%%%%%%%%%%%% ACKNOWLEDGEMENTS %%%%%%%%%%%%%%%%%%%%%%%%%%%%%%%%%
 \section*{Acknowledgements}

We warmly thank C.~Diamantini, L.~Giusti, C.~Hoelbling, 
M.~L\"uscher, K.~Rummukainen, P.~Weisz and H.~Wittig for 
useful discussions. 
This work was supported in part by the European Union
Improving Human Potential Programme 
under contracts No.\ HPRN-CT-2000-00145 ( Hadrons/Lattice QCD ) 
and HPRN-CT-2002-00311 (EURIDICE).

%-------------------------------------------------------------------

%% \newpage

\appendix
\renewcommand{\thesection}{Appendix~\Alph{section}}
\renewcommand{\thesubsection}{\Alph{section}.\arabic{subsection}}
\renewcommand{\theequation}{\Alph{section}.\arabic{equation}}

%%%%%%%%%%%%%%%%%%%%%%%%%%%%%%%%%%%%%%%%%%%%%%%%%%%%%%%%%%%%%%%%%%%%%%%%

%% \newpage

\section{Examples of quenched zero-mode integrals for $N_v > 1$}
\la{app:example}

To illustrate what can be achieved with the method discussed
in~\se\ref{nrts}, we show here how it helps to determine certain
integrals with $N_v > 1$ from known ones with $N_v = 1$. We do 
this for the flavour singlet scalar and pseudoscalar correlators, 
computed for $N_v=1$ in~\cite{DDHJ}.

Let us define 
\ba
 S^0(x) ~\equiv~ \bar{\psi}(x) I_{N_v} \psi(x)~,~~~~~~~
 P^0(x) ~\equiv~ \bar{\psi}(x) I_{N_v} i \gamma_5 \psi(x) \;.
 \label{sps}
\ea
The correlators of $S^0(x),P^0(x)$ have two parts: 
a constant and a space-time dependent contribution. 
We consider here just the former. As in~\cite{DDHJ},
we denote the constants for $S^0,P^0$ 
by $C_S^0$ and $C_P^0$, respectively. For
$N_v=1$ it was found that, to lowest order,
\ba
 C_S^0 &=& \Sigma\Sigma_{\nu}'(\mu) + \Sigma^2 + 
 \frac{\Sigma^2\nu^2}{\mu^2}\;, \la{cons1} \\
 C_P^0 &=& \frac{\Sigma\Sigma_{\nu}(\mu)}{\mu} - \frac{\Sigma^2\nu^2}{
 \mu^2} \;,\label{constants}
\ea
where $\Sigma_{\nu}(\mu)$ is the chiral condensate. 
Using now the prescription of~\se\ref{nrts} (\eq\nr{nrt}), 
we get the ``connected'' part of $C_S^0$ by 
a U($N_f$) computation, 
\ba
 [ C_S^0 ]^\rmi{quenched, $N_v=1$}_\rmi{connected} 
 &=& \lim_{N_f\to 0} \frac{1}{N_f}\left(
 \frac{\Sigma^2}{4}\Bigl\langle 
 (\tr [U_0+U^{\dagger}_0])^2 
 \Bigr\rangle \right) \cr
 &=& \lim_{N_f\to 0} \frac{1}{N_f}\left(\Sigma^2N_f\left[\frac{
 \Sigma_\nu'(\mu)}{\Sigma} + N_f\left(\frac{\Sigma_{\nu}(\mu)}{\Sigma}\right)^2
 \right]\right) 
 %% \cr &=& 
 = \Sigma\Sigma_{\nu}'(\mu) ~.
\ea
Compared with the $N_v =1$ result 
in~\eq\nr{cons1}, this gives
the full expression,  
\ba
 [ C_S^0 ]^\rmi{quenched} 
 ~=~ N_v\Sigma\Sigma_{\nu}'(\mu) + N_v^2\left(\Sigma^2 + 
\frac{\Sigma^2\nu^2}{\mu^2}\right) ~.
\ea
Similarly, for the constant $C_P^0$ we get the ``connected'' part
from a U($N_f$) computation,  
\ba
 [ C_P^0 ]^\rmi{quenched, $N_v=1$}_\rmi{connected}   
 &=& - \lim_{N_f\to 0} \frac{1}{N_f}\left(
 \frac{\Sigma^2}{4}
 \Bigl \langle(\tr [U_0-U_0^{\dagger}])^2 \Bigr\rangle \right) \cr
 &=& \lim_{N_f\to 0} \frac{1}{N_f}\left(\Sigma^2N_f\left[\frac{1}{\mu}
 \frac{\Sigma_{\nu}(\mu)}{\Sigma} - \frac{\nu^2N_f}{\mu^2}
 \right]\right) 
 %% \cr &=& 
  = \frac{\Sigma \Sigma_{\nu}(\mu)}{\mu} ~.
\ea 
Comparing this with the $N_v=1$ result, the general
expression for any $N_v$ is seen to read
\ba
 [ C_P^0]^\rmi{quenched} ~=~ \frac{N_v\Sigma\Sigma_{\nu}(\mu)}{\mu} - 
 \frac{N_v^2\Sigma^2\nu^2}{\mu^2} ~. \la{last}
\ea
\eq\nr{last} is in agreement with what one gets from
a chiral Ward Identity at fixed topological charge for $N_v = 2$, as explained 
in~\cite{DDHJ}. The space-time dependent parts of the two singlet 
correlation functions can be constructed for arbitrary $N_v$
in a similar way.

%%%%%%%%%%%%%%%%%%%%%%%%%% BIBLIOGRAPHY %%%%%%%%%%%%%%%%%%%%%%%%%%%%%%%%%%%
%%%%%%%%%%%%%%%%%%%%%%%%%%% REFERENCES %%%%%%%%%%%%%%%%%%%%%%%%%%%%%%%%%%%%

\end{document}